\def\year{2018}\relax
\documentclass[letterpaper]{article} %DO NOT CHANGE THIS
\usepackage{aaai18}  %Required
\usepackage{times}  %Required
\usepackage{helvet}  %Required
\usepackage{courier}  %Required
\usepackage{url}  %Required
\usepackage{graphicx}  %Required
\usepackage{subfig}
\usepackage{enumerate}
\begin{document}

\title{Crowdsourcing real-time viral disease and pest information.\\
A case of nation-wide cassava disease surveillance in a developing country.}
% \thanks{This research is funded by the Bill \& Melinda Gates Foundation, under the Mcrops project, at the Artificial Intelligence Research Labs}}
\author{Daniel Mutembesa$^{1}$, Christopher Omongo$^{2}$ and Ernest Mwebaze$^{1}$\\
$^{1}$Makerere University\\
P.O Box 7062, Kampala, Uganda.\\
$^{2}$National Crops Resources Research Institute\\ P.O Box 7084. Kampala, Uganda.\\
(\emph{mutembesa.daniel@gmail.com, emwebaze@cit.ac.ug, chrisomongo@gmail.com})
%Palo Alto, California 94303\\
}

\maketitle
\begin{abstract}
In most developing countries, a huge proportion of the national food basket is supported by small subsistence agricultural systems. A major challenge to these systems is disease and pest attacks which have a devastating effect on the small-holder farmers that depend on these systems for their livelihoods. A key component of any proposed solution is a good disease surveillance network. However, current surveillance efforts are unable to provide sufficient data for monitoring such phenomena over a vast geographic area efficiently and effectively due to limited resources, both human and financial. Crowdsourcing with farmer crowds that have access to mobile phones offers a viable option to provide all year round real-time surveillance data on viral disease and pest incidence and severity. This work presents a mobile ad hoc surveillance system for monitoring viral diseases and pests  in cassava. We present results from a pilot in Uganda where this system was deployed for 76 weeks. We discuss the participation behaviours of the crowds with mobile smartphones as well as the effects of several incentives applied.

%Agricultural subsistence systems by small-holder farmers support 90\% of the national food basket, are threatened by viral disease outbreaks and pest attacks, which cause massive yield loss and food insecurity in Uganda. Current expert survey efforts are unable to provide sufficient surveillance data for monitoring such phenomena over a vast geographic area efficiently and effectively within limited resources. Crowdsourcing with farmer crowds that have access to mobile phones, offers a viable option to harness all year round real-time surveillance data on viral disease and pest incidence and severity for Uganda. This work presents the mobile ad hoc surveillance system and its pilot implementation for monitoring viral disease and pest using real-time cassava surveillance data crowdsourced from farmers and experts in-field all year round across Uganda. The research presents insights into the participation behaviours of the crowds with mobile smartphones and under the influence of several incentives.
\end{abstract}

\section{Introduction}
Real-time surveillance forms the basis for effective crop health monitoring and disease detection. In Uganda, where viral disease attacks on crops are one of the leading causes of food insecurity and poverty, having a functional surveillance system is critical. In this work, we focus on the cassava crop (Manihot esculenta), an important food security crop for Sub-Saharan Africa and other regions. Cassava is the second most important food and security crop in Sub-Saharan Africa especially amongst smallholder farmers because it can easily be grown in poor soils and requires few inputs. Although cassava is known to survive under harsh conditions, its yield is greatly affected by pests and viral diseases \cite{ephraim2015,otim2000current}. 

%Cassava (mahot blah blah) Viral diseases amongst staple food crops such as Cassava in Uganda and the region are reason for reduced yields in subsistence agriculture, particularly among smallholder farmers (Nape et al. 2015).

Currently, surveillance is done by experts from the national agricultural organisation in Uganda, who will every year conduct a national cassava disease and pest survey. This generally entails inspection of gardens at specific intervals (of about 10Km) along drivable main roads, covering only small proportions of the areas of interest in major districts. Because they have to work within limited budgets, this survey may be delayed or fewer regions sampled in a particular year. These expert surveys whilst providing an annual snapshot of the health of cassava across the country, are limited in their ability to provide real-time actionable surveillance data. 

In many other fields, citizen science has proven to be a good intervention to supplement long-term focused systems such as the surveillance system employed in Uganda \cite{silvertown2009new}. Citizen science and more specifically crowdsourcing can be employed to supplement such a system by providing ad hoc data input from citizens or crowds nearer to the phenomenon of interest. In the Ugandan case, small-holder farmers and extension service workers who are on ground in the various regions of the country can provide supplemental surveillance data more frequently and more efficiently using smartphones.

A key facilitator to this process is mobile telephony. The ubiquity of mobile phones in developing countries presents a unique opportunity to leverage crowdsourcing for this type of problem, and several example implementations of this already exist \cite{chatzimilioudis2012crowdsourcing,agapie2015crowdsourcing}. Mobile crop surveillance in areas where it is resource infeasible to survey regularly offers a viable option for effective crop sensing, enabling experts to conduct surveillance tasks more effectively while at the same time crowdsourcing surveillance reports from farmers local to these areas to inform interventions in a timely manner.

This paper presents the mobile ad hoc surveillance system, which uses a crowdsourcing setup for farmer and expert crowds with mobile smartphones for monitoring of cassava viral diseases and pests in Uganda. The sections that follow describe the related work in this domain, the implementation of the crowdsourcing system used, the incentives employed in this setup and some lessons and experiences from running this set up for 76 weeks in Uganda. 

%This paper is divided into six sections; Section one, introduces our research work. Section two, presents related work, section three presents the mobile ad hoc surveillance system, its taxonomy. Section four, sets the research objective and the incentives applied in the pilot. In section five, we present results from  of our approach, crowdsourcing under incentives. In section six, we state our contribution and make recommendations on our approaches.

\section{Related Work}
In this age of big data, there is an ever increasing need to build applications whose strength lies to a significant degree on the amount of data used to train the system. For such system it quickly becomes infeasible for a small team of experts to collect all the data. Citizen science and crowdsourcing applications are thus becoming increasingly important in accomplishing this task \cite{kittur2013future}, some notable early examples in the field including Open Mind Common Sense (OMCS) \cite{singh2002open} and LEARNER \cite{chklovski2003learner} where knowledge is collected from a crowd of minimally skilled volunteers. More recently we have Amazon Mechnanical Turk \cite{paolacci2010running} and Crowdflower \cite{van2012designing} as examples of popular large scale crowdsourcing platforms.

This work is broadly enshrined under the Knowledge Collection from Volunteer Contributors (KCVC) \cite{chklovski2005towards} which is more aptly enshrined under the broad themes of crowdsourcing and collective intelligence \cite{quinn2011human}. A significant component of the work falls under the human computation field as well since the contributors in this system have to provide some analysis of the image they are uploading. Because we presently do not evaluate how accurate these particular analyses are, the work we present here mainly focuses on the crowdsourcing aspect. We see several parallels in our work with work done on organizing the crowd in focused groups \cite{zhu2012organizing} and the more related discipline of voluteered geographical information (VGI) \cite{haklay2013citizen}. Several examples however of applying crowdsourcing with incentives for human computation tasks exist including \cite{kanefsky2001can} where the crowd was used to classify craters from images of Mars.

A key component of this work is the ability to leverage the ubiquity of smartphone technology particularly in the African context. Mobile phone sensing offers a variety of novel, efficient ways to opportunistically collect data, enabling numerous mobile crowdsourced sensing (MCS) applications. Examples from alternate fields include \cite{mohan2008nericell} and \cite{thiagarajan2009vtrack} where systems are developed for crowdsourcing traffic information. More related examples to our cassava disease sensing application include the use of crowds and communities in noise pollution sensing \cite{rana2010ear} and in air pollution sensing \cite{stevens2010crowdsourcing}. An even closer example of a system using different incentives to link credible buyers to sellers of market produce in Uganda is the Kudu system \cite{ssekibuule2013mobile} which uses a double auction design algorithm to link buyers and sellers of agricultural produce. 
% For more details, we refer interested readers to several survey papers [1, 8, 9]. Mobile phones are massively available in East Africa and Africa generally.
\\
% Kudu [61] is a double auction agricultural market platform that links willing buyer-willing seller efficiently by voluntary participants who are farmers and agro produce dealers in the market. Over a mobile based network, the farmers post their location, quantities of produce ready for market and the their asking price for the goods using a mobile phone. The market buyers also post the goods they are interested in, their location and their giving price, also known as their bid, using their mobile phone.
% The double auction algorithm then matches them based on these offers and bids, and successful matches receive notification and may proceed to negotiate in person. This method could be harnessed for crowdsourcing agricultural market trends in production and market demand.

%%%%%%%%%%%%%%%%%%

% \begin{figure}
% \centering
% \includegraphics[width=8cm]{images/news_montage2.png}
% \caption{News reports on the epidermic threat by Cassava Mosaic and Brown Streak viral diseases in the East African region.} 
% \label{Fig. 1}
% \end{figure}

%  \begin{figure}
% \centering
% \includegraphics[width=8cm]{images/Googlemap}    % The printed column width is 8.4 cm.
% \caption{A map of data collected by trailing teams of cassava experts on their once a year annual national pest and disease survey in November 2014. The experts survey major townships and only along specific intervals on main roads.} 
% \label{Fig. 2}
% \end{figure}

\section{Mobile Ad hoc Surveillance system (\emph{AdSurv}) }

For this particular project, we implemented a crowdsourcing mobile phone based ad hoc surveillance system (AdSurv) that enables farmers, extension workers and agricultural experts provide near real-time geo-tagged surveillance data for monitoring cassava crop health across Uganda. The data collected is in the form of geo-coded images of plants in gardens in the locality of the collector. These reports update a national cassava situation map. 

The current operation of the system involves the crowdsourcer broadcasting a surveillance task of a subject of interest to the agents. This is done through the crowdsourcing application on the mobile phone that the collector is using to transmit the data. Based on this probe from the crowdsourcer, the collector sends in images and text data pertaining to the request. A typical request could be "\emph{this week send images of cassava crops in your vicinity exhibiting Cassava Mosaic Disease}". The response to this would be the farmer or extension worker visiting different fields in his locality and sending images of plants exhibiting the disease. 

Since we follow a typical Knowledge Collection from Volunteer Contributors (KCVC) methodology \cite{chklovski2005towards}, we compensate the collectors with mobile phone data credit, micro payments or other types of incentives. In later sections we discuss these incentives in greater detail. For this work we carried out a pilot spanning 76 weeks with several farmers and extension workers in Uganda. The initial goal was to use a KCVC model to crowdsource image data to aid in the real-time surveillance of disease  as well as to inform the development of algorithms to automatically classify disease. The pilot was not designed initially to study the effect of different incentives though this became very important as the pilot went on. As such our assertions in this paper are mainly on the conjecture side of the line, though we believe the paper presents strong evidence to support some known theories on crowdsourcing, particularly seeing as this was applied to solve a real problem. 
% The agents then complete the surveillance task one form at a time and send it back to the server, as it is mapped in real-time. 

\subsection{The pilot}
The \emph{Adsurv} crowdsourcing system was piloted in several regions of Uganda to provide evidence of the utility of a crowdsourced surveillance data to inform actionable interventions. The contributors in the pilot were selected to include subject matter experts, extension personnel and cassava farmers from across the country. Three types of image data were identified by the national agricultural organisation in Uganda as being of priority for this system. These included images of cassava plants manifesting disease, images of cassava pests e.g. whiteflies and finally images of anomalous manifestations on cassava plants in the gardens. The anomalous images were particularly important for providing early warning signals about a potential new disease outbreak. The system was set up to upload the data to a centralized server. 

In total the pilot included 29 participants from the different regions of Uganda. Figure \ref{reports}a shows the map of Uganda and the corresponding contributions from the different regions. As is evident there was a relatively good spatial coverage of the country. A key reason for this as well is that the cassava experts in their ordinary duties also traverse the different regions of Uganda and during these trips, they were also able to send in image data from these various locations. This improved the coverage of the system. 

In week 1 of the pilot, the 29 participants were given a one day training which mainly involved training on the smartphones distributed to them, demonstration of the system, a field test of the system. Protocols and modalities of what data and how often to send the data were discussed. A second refresher training took place in week 64 of the pilot were the participants were trained on new features of the application including a better notification system for notifying them of the weeks data collection target. At this training, 15 best performing participants were in invited (as an incentive).
% Another retraining and retooling exercise was carried out at 64 weeks into the pilot. This was aimed at updating the mobile android app for AdSurv and resolving technical issues for some of the agents.

\subsection{The \emph{AdSurv} crowdsourcing eco-system}
The \emph{AdSurv} crowdsourcing system was made up of four key entities including (i) the crowd consisting of the contributors, (ii) the crowdsourcer who seeks the knowledge and wisdom of the crowd, (iii) the crowdsourcing task which is the activity in which the crowd participates and (iv) the crowdsourcing platform which is the system that uses the mobile phone network to make communication between the crowd and the crowdsourcer possible. We expound on these four entities highlighting how we instantiated them for the pilot.
% 1. The crowd, which consists of the people who take part in a crowdsourcing activity.\\
% 2. The crowdsourcer, is the entity (a person, a for-profit organisation, a non-profit organisation, etc.) who seeks the power and wisdom of the crowd for a task at hand.\\
% 3. The crowdsourcing task, also simply called the task, is the activity in which the crowd participates.\\
% 4. The crowdsourcing platform, is the system within which a crowdsourcing task is performed.

\subsubsection{The Crowd:}
The \emph{AdSurv} system was deployed countrywide by availing mobile smartphones with the application to experts, extension workers and farmers, located across 4 of the target 6 agricultural sub-regions of country. A total of 29 persons were equipped with a smartphone.  The smartphones were pre-installed with the AdSurv crowdsourcing app and the volunteers were trained on how to use the the app for collecting and transmitting data. The over-arching goal was to evaluate our hypothesis that a crowd consisting of small-holder farmers and extension experts can supply near real-time surveillance data particularly for times when information about the same is lacking for example in the periods between the standard national surveillance surveys.

For this pilot, the crowd consisted of four categories of participants or agents, all with varying expertise and profiles. These included farmers, crop experts, extension service members, and the partner agents to the National Crops Resources Research Institute (NaCRRI) the Uganda government body that is in charge of cassava crop disease surveillance. A concise description of each group follows.
\begin{enumerate}
\item Farmers - these were small-holder farmers in the four major agro-ecological zones of the country that have communities growing cassava.
\item Experts - these were cassava crop subject matter experts who conduct survey related tasks as one of their research duties.
\item Extension service workers - these included district based officers for example district agricultural inspectors or district production officers whose work it is to routinely visit and inspect farmer fields and to disseminate information on farming best practices.
\item Partner agents - these were non-governmental extension service support workers who support the agricultural sector in Uganda. Their roles and duties are largely similar to the extension service.
\end{enumerate}

\subsubsection{The Crowdsourcer:}
The crowdsourcer also known as the principal, is the institution that requests the data or can make use of the data. In our case, the cassava disease surveillance task is the mandate of the the Cassava Regional Centre of Excellence at the National Crops Resources Research Institute (NaCRRI). Together with the Artificial Intelligence and Data Science Research Lab at Makerere University, these formed the crowdsourcer. The role of the crowdsourcer is to provide the technical equipment, incentives for participation, and define the survey tasks to be completed by the crowd.

\subsubsection{The crowdsourcing task:} 
In the case of our pilot, the crowdsourcer publishes weekly crowdsourcing tasks on surveillance subjects of interest. For the pilot, the surveillance subjects mainly included tasks to collect images of plants depicting cassava viral diseases like Cassava Mosaic Disease, or pests like Cassava African Whitefly and Cassava Green Mite, or disease symptoms e.g. necrotised root sections of cassava. Other tasks included submission of statistics on numbers of cassava farmers in a village and cassava yield, among others.

Every report related to a particular crowdsourcing task included four attributes; (i) the image of the subject of interest, e.g. a diseased leaf, (ii) textual data specifying the type of image which could be a disease manifesting image, or a pest image, or an anomalous image, (iii) a comment from the participant or agent highlighting their understanding of the image they are uploading, and (iv) the GPS reading of the area where the image was taken. A complete report is one with all these four attributes accounted for.

% First attribute of the task is taking an image of the subject, for example cassava leaf or cassava root tuber, and filling in form on the attributes of the image taken.
% For the data collection form, the second attribute is the subject image, then image-type, which categorises the subject of that image e.g disease image, whitefly vector image, an anomaly, other, for images that are not related to disease and vector surveillance.\\
% The third attribute is the comment/observation section which allows for the reporter to fill in the variety of cassava and the observation made on the crop for the image taken.\\
% The forth attribute is recording geo-coordinates (GPS) for that particular image or report.\\
% Every report has these four attributes and is considered complete when all have been filled in.

\begin{figure}
\centering
\includegraphics[width=0.48\textwidth]{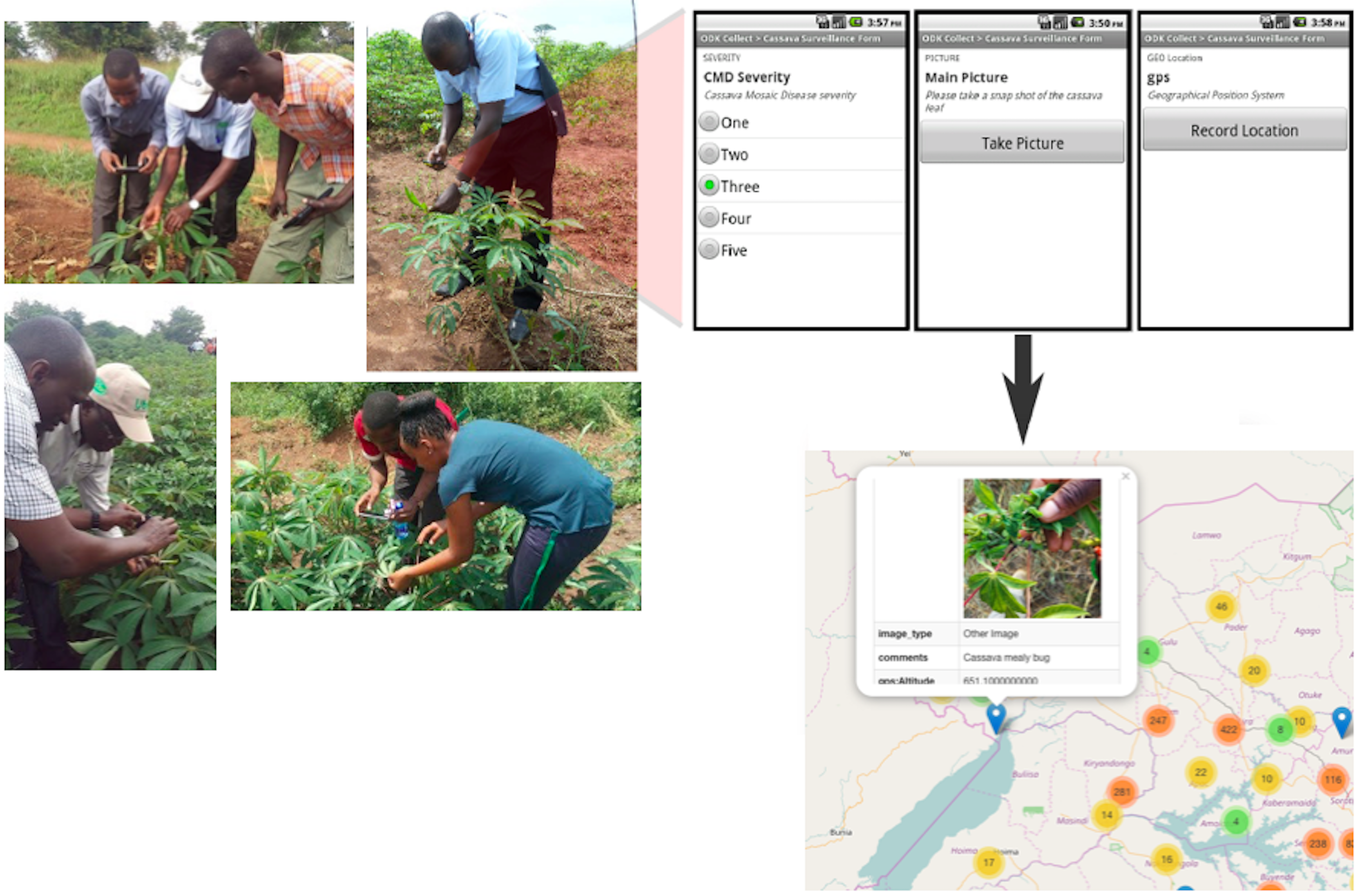}
\caption{The \emph{AdSurv} crowdsourcing system.}
\label{adsurvsystem}
\end{figure}

% \begin{figure}
%  \centering
% \includegraphics[width=8.5cm]{images/AdSurv_pipeline}

% \caption{The ad hoc surveillance system pipeline; mobile app, crowdsourcing task, online real-time map populated with agent reports.}
%  \label{Fig. 4}
% \end{figure}

% \begin{figure}
% \centering
% \includegraphics[width=7cm]{images/AdSurv_task}    
% \caption{The Surveillance task on the AdSurv app on smartphone, showing the steps of how to fill in an AdSurv form; take a picture, assign a label to the report/image, comment or observation, record GPS, upload to server } 
% \label{Fig. 3}
% \end{figure}

\subsubsection{The crowdsourcing platform:}
This consists the actual mobile ad hoc surveillance system which includes, the smartphone application, the backend server and a web interface to the server system. The next section describes the system components in detail.

\subsection{AdSurv system architecture}
The Adhoc surveillance system consists three components that ensure its robust functionality; the AdSurv Android app, back-end server and a web interface to allow interaction with the system. Figure \ref{adsurvsystem} gives a depiction of the system.

\paragraph{AdSurv Android app:} 
The AdSurv platform consists of a smartphone based Android application, called Adsurv, used for the data collection activity. The AdSurv app builds upon the Open Data Kit (ODK)\footnote{https://opendatakit.org}, an open source Android data collection application. The mobile app has data collection forms which are used for capturing details about the subject of interest. These include the image of the plant, a label classifying the type of image, and a GPS reading of the location from which the data is being collected. The data is uploaded to a centralized server and automatically mapped in real-time.

\paragraph{AdSurv back-end server:}
At the back-end, the system uses a customized database application that receives data from the Android application and runs some summary statistics on the data to understand the number of uploads per farmer, the trend of submissions, etc. The data is also mapped as well as posted to a dashboard accessible to the users of the system. 

\paragraph{The web interface: } 
The web interface\footnote{https://adsurv.mcrops.org} to the application consists of a mapping application and a dash board. The mapping application uses Google Maps to populate the mobile phone uploads to a live real-time map. The mapping application is interactive and can be used to generate a heat map version of the density of disease over the whole country. The dash board presents basic statistical summaries of the upload report features, the data collected and the reporting trends. 

% The dashboard tab 1 shows summaries of the kind of image types that have been reported visually by a bar graph.\\ 
% Tab 2 shows a bar chart of individual contributions totals for a given period.\\
% Tab 3 shows a visual representation of individual contributions per category of image or report type.\\
% Tab 4 shows a time series bar graph that plots totals monthly over a given period.

% \begin{figure}
%  \centering
 
% \includegraphics[width=8cm]{images/AdSurv_Dashboard}

% \caption{The ad hoc surveillance dashboard.}
%  \label{Fig. 5}
% \end{figure}

\subsection{Incentives structure}
For crowdsourcing, the incentive structure is critical. Whilst the pilot was not built to test what incentives work, we invariably ended up trying different incentives and comparing the performance of the contributors based on the incentives. In total we applied six different types of incentives over the lifetime of the pilot. These were applied in a semi-consecutive, semi-mixed fashion based on the numbers of contributions that were coming in. These included the following. 
\begin{enumerate}
\item Provision of equipment - at the onset of the pilot, we provided the participants with \$100 smartphones with the AdSurv app installed. Participants were required to sign a contract that had as one of its provisions, continual ownership of the equipment being dependent on performance in the crowdsourcing task. 
\item Provision of mobile data credit - this incentive was mainly to facilitate the sending of reports to the central server. We iterated through different implementations of this. We tried sending the data credit before upload of the reports, we tried sending the credit on submission of the reports and we tried monthly versus weekly credit disbursements. 
\item Weekly and monthly prompts - we maintained a weekly communication by text SMS to every participant and a monthly telephone follow up call. These were mainly reminder alerts. These were initially quite effective.
\item Provision of a focused subject for the week - we termed this \emph{subject surveillance} to imply the participants were requested to report on very specific phenomena every week. This was in response to feedback from the participants sending the \emph{same thing} every time.
\item Provision of feedback - here we provided the participants with some form of analysis from the images they had uploaded.
\item Provision of micropayments - at some point in the pilot, we started providing direct monetary rewards with a \emph{pay-per-report} model. We varied the amounts of micropayment during the pilot resulting in different responses from the participants.
\end{enumerate}

% \subsection{Incentives schedule}
Incentives were mostly applied in a relatively ad hoc manner, without a predefined schedule. They were mainly in response to feedback from participants explaining varying trends in the reports uploaded to the system. The incentives were also combined randomly from time to time over the pilot period of 76 weeks. 

The incentives issued to motivate participation and reporting were generally allowed to run for as long as the agents were reporting above the desired compliance threshold level. Compliance threshold was set to 10 reports per week on a particular task. Some incentives were sustained longer than others.
%The figure below show the incentives map for the pilot implementation of the mobile ad hoc surveillance.

%\begin{figure}
% \centering
 
%\includegraphics[width=8cm]{images/AdSurv_Incentives_map}

%\caption{Incentives applied on the ad hoc surveillance volunteer agents for a period of 76 weeks.}
% \label{Fig. 5}
%\end{figure}

\section{Results}
The pilot was set up to collect near real-time surveillance data on cassava diseases in Uganda. To this end, more than 7,000 reports was collected over a period of 76 weeks. This is about a third of the expected number of reports from a group of 29 participants over that same period. 

\begin{figure*}[!h]
  \centering
      \subfloat[Spatial submission of reports]{
      \includegraphics[width=.45\textwidth]{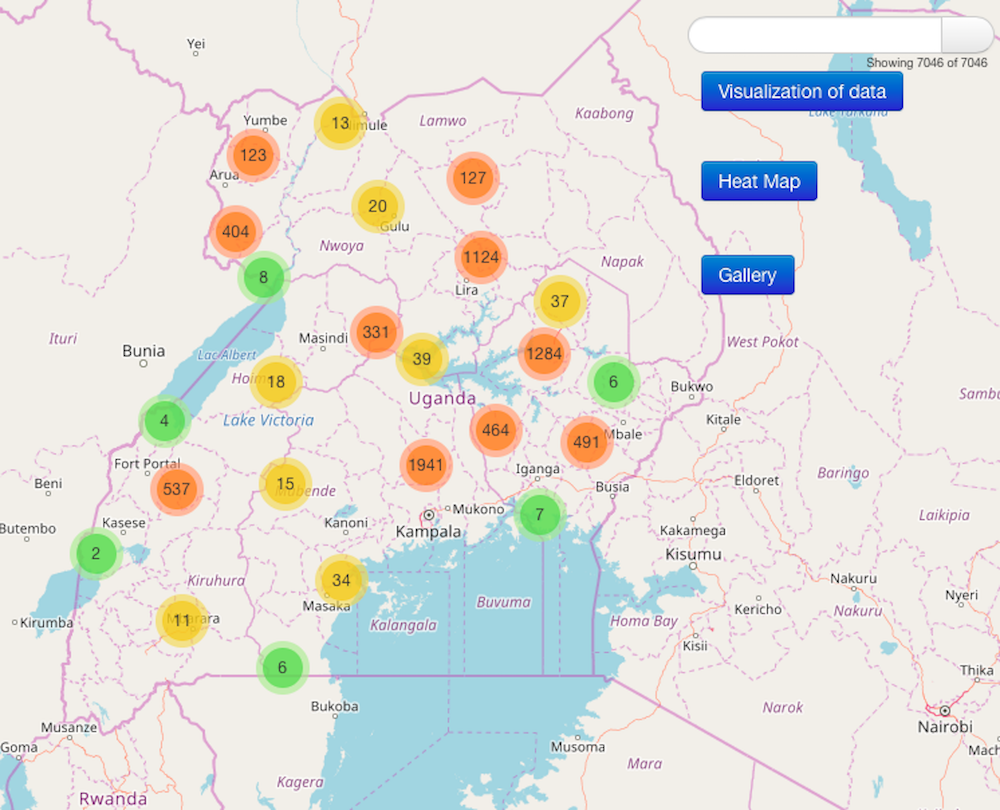}}
    \subfloat[Trend of report submission]{
      \includegraphics[width=.45\textwidth]{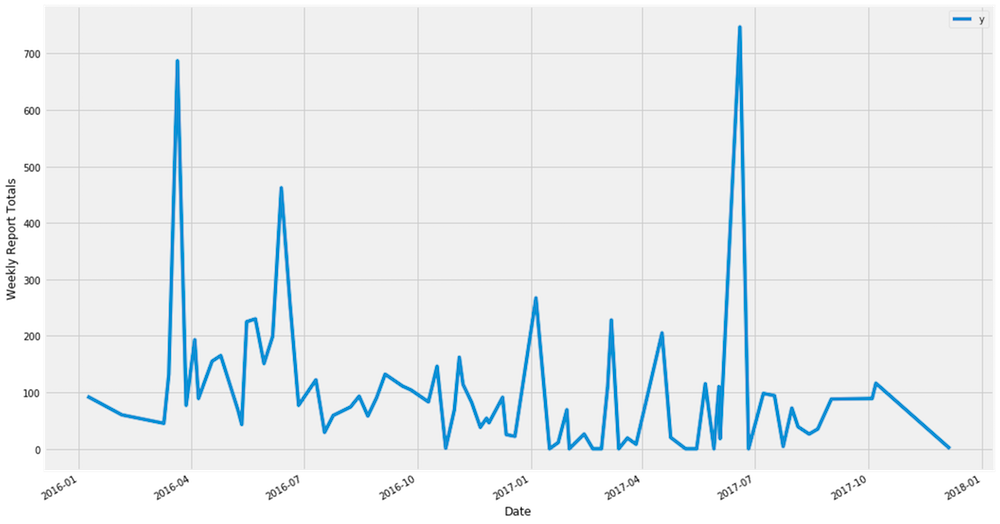}}
    \caption{Reports collected using \emph{AdSurv} system; spatial and timeseries dimensions.}
    \label{reports}
\end{figure*}

One plausible explanation for this number is that we had some of the participants drop out of the pilot, when equipment was stolen or when the equipment did not work as was the case with GPS on some of the devices. GPS was a mandatory field on the form so participants were unable to send in reports if they were not geo-coded. Figure \ref{reports}a depicts the reports collected over the different regions of Uganda. Figure \ref{reports}b shows the corresponding timeseries of the report uploads over the 76 weeks of the pilot. The next sections explain the trends observed in these figures.

% \begin{figure}
%  \centering
 
% \includegraphics[width=8cm]{images/AdSurv_Reports_76weeks}

% \caption{The plot shows the reporting trends of agents for a period of 76 weeks}
%  \label{Fig. 6}
% \end{figure}

% \begin{figure}
%  \centering
 
% \includegraphics[width=8cm]{images/AdSurv_Map1}

% \caption{A map of Uganda showing data points of geo-coded surveillance reports submitted by agents to AdSurv from all across the country for a period of 76 weeks.}
%  \label{Fig. 7}
% \end{figure}

Most of the reports were of images depicting disease, followed by other types of images including farmer images, images of roots harvested etc. Images of whitefly infested leaves and anomalous images were the lowest recorded of the 7000. This is because the crop experts who were the biggest contributing group were mostly interested in disease related data from all around the country. This was also influenced the nature of subject surveillance calls made during the pilot.

While the system was setup to collect surveillance data, we eventually got involved in understanding how the different incentives affected the participation of the crowd. The next two sections present results from these other dimensions of the pilot.

\subsection{Reporting by agent category}
\begin{figure*}[!h]
  \centering
      \subfloat[Example submissions per category]{
      \includegraphics[width=.45\textwidth]{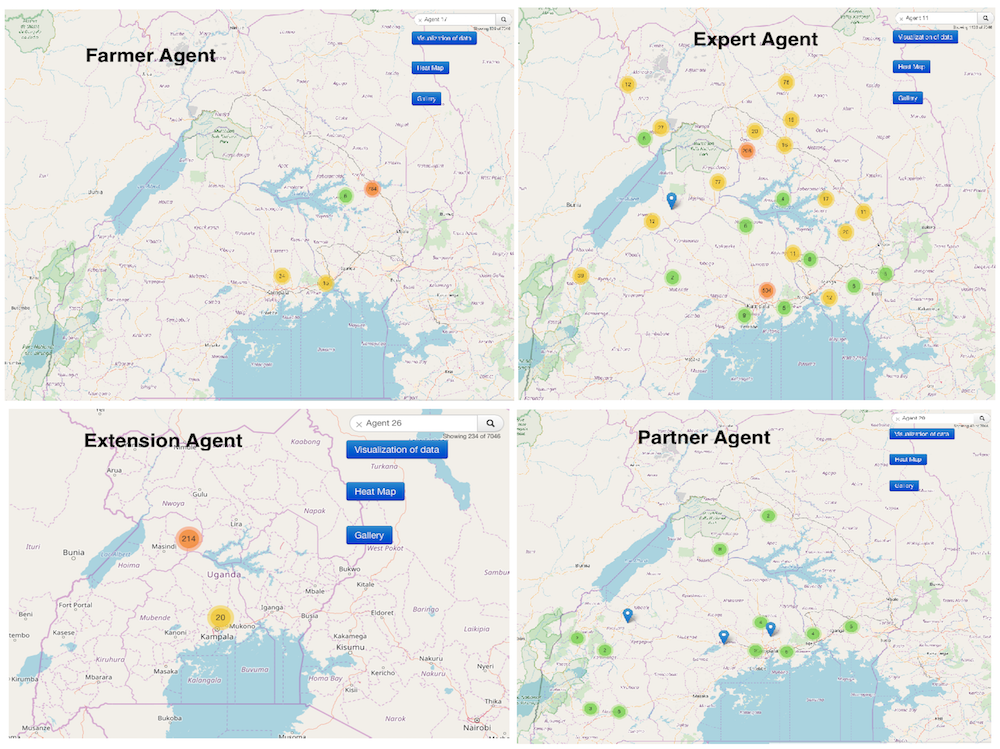}}
    \subfloat[Trends per category]{
      \includegraphics[width=.45\textwidth]{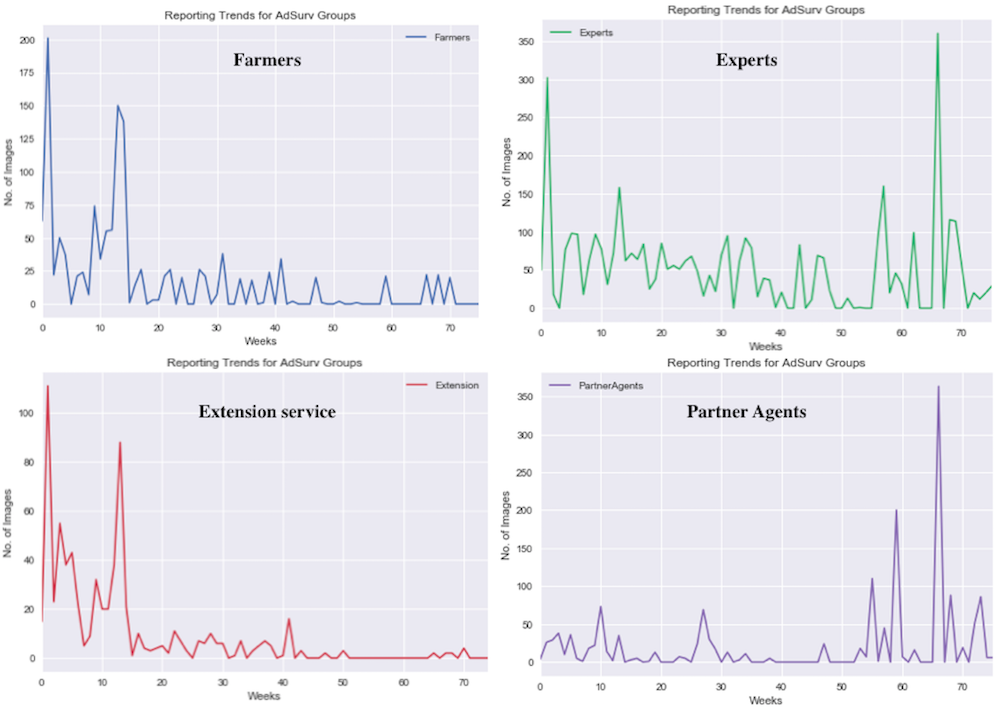}}
    \caption{Reports collected using \emph{AdSurv} system; depiction of the different user categories.}
    \label{categoryreports}
\end{figure*}

Figures \ref{categoryreports}a and \ref{categoryreports}b present a depiction of the contributions of the different types of agents in the pilot. Figure \ref{categoryreports}a provides single examples of individuals from the different participant categories while Figure \ref{categoryreports}b provides the overall trend of reporting for all participants across the different categories. As is evident the experts who travel around the country more often have a wider spatial spread of contributions compared to the other categories of participants. In general, our findings were that each of the volunteer categories in the pilot exhibited unique participation behaviours, and were motivated differently by different incentives. More specifically;
\begin{enumerate}[-]
\item \emph{Farmer agents} tended to report strictly from their localities or regions for the entire period of the pilot. Even for very active and consistent farmer agents like farmer agent 17 (Figure \ref{categoryreports}a), with a total submission of over 700 reports, most of the reports are submitted from a very specific location. We found that these farmer agents are more influenced by non-monetary incentives especially communication type incentives like follow up calls and SMS message alerts. They also mostly reacted positively to reputation type incentives.

\item \emph{Extension service agents} are also closer to the farmer agents in terms of their spatial spread. Figure \ref{categoryreports}a depicts the report contribution from a typical extension service agent. The limited spatial spread is probably because extension service members, such as district agricultural crop inspectors and district production officers, are localised in their operations and assignments. Overall we found these to be the least performing and the least influenced by direct monetary type rewards and yet their nature of work gives them access to many farmers fields on a regular basis. 

\item For \emph{expert agents}, we notice significant spatial spread of the collections. An example of a typical expert agents collection is depicted in Figure \ref{categoryreports}a. Many of the expert agents reported from areas other than their research station localities, many times covering disparate regions of the country. We found that expert agents are most influenced by higher direct monetary reward incentives and they generally posted higher quality reports especially on the subject surveillance matters probably because of the specialised knowledge they possess. The expert agents generally reported from disparate areas across the country possibly while they were carrying out their routine field activities. 

\item For the \emph{partner agents}, we generally noticed a pattern between that exhibited by the experts and that exhibited by the farmers - some localisation in the collection yet at the same time some leanings towards sparsity. The partner agents were particularly most influenced by the introduction of direct monetary rewards. This is a behaviour that is very interesting for further research.
\end{enumerate}

\subsection{Reporting trends by category}
Figure \ref{categoryreports}b depicts the reporting trends for the different categories of participants or agents in the crowd over the 76 week period of the pilot. We make specific observations based on this.
\begin{enumerate}[-]
\item For the first twenty weeks, mobile phone data credit/airtime facilitation was provided to the agents to report on the surveillance tasks. Several follow up calls were also made to the agents, explaining the high peaks in the first few weeks. However, we notice a steady decline in reporting over the twenty weeks with the two major incentives. From the 20th to the 30th week, there are several weeks where the agents were not provided weekly data airtime, but a constant stream of 2 SMS messages per week were broadcast to them. The participants over this period generally maintained a compliant level of reporting.
\item Subject surveillance is where a specific subject of interest to the crowdsourcer is broadcast to the agents as a surveillance task for each week.
This was introduced in the 30th week, which led to the slight rise in reporting in the 31st week. This however, gradually decays over 10 weeks, even with a steady facilitation with weekly data airtime to support the subject surveillance tasks. The subject surveillance tasks were doubled in the 37th week but do not register any success in influencing the reporting upwards. Reporting for this period continued to steadily decline even with weekly data airtime facilitation provided.
\item In the 42nd week we introduced a direct monetary reward incentive where agents were directly rewarded based on the number of reports submitted. Micropayments were done on a per report basis. The first reward amount was 250 Ugandan shillings (approx. 5 cts USD) per report which run for 12 weeks from the 42nd to the 54th week. This was only able to slightly affect the reporting trends in the 43rd, 45th and 46th weeks combined with other incentives of weekly data airtime provided to agents and subject surveillance task being broadcast every week.
\item In the 55th week, the direct monetary reward was increased from 250 Ugandan shillings to 500 Ugandan shillings per report, and we immediately saw a steady rise in reporting trends from the agents from the 55th to the 60th week of the pilot. This was coupled with follow up calls and SMS messages to let the agents know of the new incentive of direct monetary rewards for reports sent in, and a broadcast of two subject surveillance tasks in the same period of time. This caused 3 peaks of reports in the 57th, 59th and 63rd weeks with majority of the reports coming from the experts and the partner agents.
\item In the 56th week, the weekly data airtime facilitation was withdrawn immediately after the increase of direct monetary reward which was structured to be paid after an agent submitted reports. This seemed not to have a positive effect on the farmers and extension service agents.
\item The great peak in week 68, was largely due to a retraining and retooling exercise we held for the agents who had adhered to compliance of posting on average 100 reports for at least 60 weeks from the start of the pilot. After the retraining and retooling seminar, we see a steep rise in agents contributions for weeks of 67-68 but which then steeply drops to slightly above 200 reports per week. At this point only two groups of agents are active i.e. experts and partner agents.
\item For the reminder of the pilot, we applied a direct monetary reward for reports, subject surveillance coupled with weekly SMS message alerts to agents. These were most exploited by the expert and partner agents. We see a gradual decline in overall reporting trends still from the 70th week onwards.
\end{enumerate}

% \begin{figure}
%  \centering
 
% \includegraphics[width=8.5cm]{images/AdSurvGroupsReportTrends}

% \caption{The plot shows the reporting trends for agent groups reporting to ad hoc surveillance.}
%  \label{Fig. 8}
% \end{figure}

% \paragraph{AdSurv groups}

% \begin{figure}
%  \centering
 
% \includegraphics[width=8cm]{images/ClusterReport_Behaviour}

% \caption{The maps show the clustered reporting behaviour expressed by agents of the farmer and extension service crowds. 
% Top left-to-right clockwise are maps of; farmer agent, an expert agent, partner agent and extension service agent.}
%  \label{Fig. 9}
% \end{figure}

\subsection{Effect of incentives}
In this pilot, the effect of incentives on the reporting trends generally varies with the type of incentive or more precisely with the type of incentives mixed at any one point. Generally for all bundles of incentives, we notice a positive effect at the onset of the incentive and then a gradual decline to some steady state. The different agent categories are also influenced differently by the different incentives. More generally we observe the following.
\begin{enumerate}
\item Incentives played a crucial part in motivating reporting from autonomous agents in disparate areas across the country for the period of 76 weeks.
\item All the agents were strongly motivated by the mobile smartphones that were given to them for the surveillance task at the start of the pilot program. This lasted a couple of weeks.
\item Farmers seem to be most motivated by the non-monetary incentives like communication-type incentives. We observed that they reported more after a followup call was made which was done about once every four weeks. During the first 20 weeks of the pilot, reputation type rewards were used where we broadcast the best performers of the week. This was a big motivation for farmer agents and yet did not influence much the experts and partner agents.
\item We found that the partner agents were the least influenced of all the categories by facilitation incentives and non-monetary rewards like communication incentives of SMS messages, subject surveillance and follow up calls. 
\item The direct monetary rewards of \emph{pay-per-report} most incentivised the crop experts and the partner agents and were not able to positively influence the reporting from farmer and extension service agents.
\item We noticed that the higher the direct monetary reward, the higher the participation of experts and to a small extent, partner agents. With a direct monetary reward, experts and partner agents were more motivated to report. 
\end{enumerate}

Each bundle of incentives were able to generally sustain a reporting trend for on average 10 weeks. With such incentives applied periodically and interchangeably for every 10 weeks on average, it seems possible to incentivise this particular crowd for desired participation and reporting.

\subsection{Discussion on challenges}

During the pilot, we encountered several challenges, some technical, and some social. Many of the interventions and incentive changes during the pilot were a response to one of these challenges. For some we had no immediate solution or the solution was too expensive (in terms of resources and time) for us to implement in the pilot. The challenges generally included:
\begin{enumerate}
\item Most of the farmer and extension service agents were not very conversant with using mobile smartphone technologies. They largely have and use  feature phones. This put a direct strain on the training budgets and most of the problems faced during the period of 76 weeks were concerned with using the mobile AdSurv app on the smartphone. However, over time the agents were able to operate some social media mobile apps, navigate the phonebook, use the camera, make calls and message with the mobile smartphone with relatively much more ease. Perhaps future research efforts may investigate a gamified crowdsourcing smartphone app, and increasing the utility features of the app to include for example commodity market information, news channels for high yielding plant materials etc.
\item When provided with weekly data facilitation, some of the agents used the data bundles for their own purposes like chatting on social media apps, and exhausted it well before being able to submit any reports. This behaviour was most dominant when weekly data facilitation was sent to them at the beginning of the week for example on Monday and Tuesday before they had to upload reports on Friday. After a follow up call and survey in the 9th week, it was noticed that most of the agents preferred submitting their weekly collections at the end of the week on days like Friday and Saturday. Starting with week 11, data facilitation was posted to the agents weekly on Fridays. This improved submissions somewhat.
\item We had many challenges related to the agents accessing mobile telephone internet with their handsets. This resulted into demotivation for some of the agents after several failed attempts at submitting reports. We got several complaints of non payments for reports which had supposedly been sent. It was unclear whether they were actually sent and lost during transmission or otherwise. From time to time, we advised the agents to collect the data and find township areas that have a good mobile internet coverage so as to send the reports uninterrupted. This however, was not as forth coming especially for agents located in rural places.
\item There was a significant problem with agents being unable to send the reports because the GPS coordinates on the app could not resolve or because they took more than 15 minutes to resolve on the phone. It was unclear whether this was due to poor connectivity or due to low quality GPS sensors in the mobile devices. Because this was a required field for the report, this quickly became a point of frustration for the agents in those areas, many times demotivating some of the agents leading to a poor and irregular performance. We made several technical visits to these places to ascertain the cause and for most cases we found the experience of getting GPS coordinates was as described (taking more than 15 min). For the most part, this was a mobile device dependent problem.
\item Some of the agents lost, broke and were robbed of their equipment which reduced the number of contributing agents from 29 to 27 agents in total. There was no way to know if the stories of lost phones were true or not. 
\item Many of the agents from the rural areas and the older volunteers (above 60 years of age) were not conversant with usage of mobile smart phone technology. In the training sessions, we found that they comprehended better when guided by their fellow farmers, mostly in a local dialect.
Future training will take into consideration decentralised training programs customised to the language requirements of that particular region, delivered by experienced model farmers.
\item With the increase of the direct monetary reward we experienced scenaries where an agent would report many reports from within a very small restricted locality. This was especially so with the partner agents. We observed numbers of reports for particular agents and experts shoot up from 12 to over 70 per week when the micropayment incentive was increased to 500 UGX per report. This was later countered by putting a limit on how many reports would be paid for from the same village per week. The limit was set to a maximum of 35 rewardable reports.
\end{enumerate}

\subsection{Recommendations from the pilot}
The pilot of the mobile ad hoc surveillance system brought to light some pertinent issues in working with voluntary participants from rural areas. Some unaccounted for factors like the different levels of expertise, different literacy levels were found to affect the quality of data collected and the frequency of reporting. Some recommendations on what key factors to consider in deploying such a system include:
\begin{enumerate}
\item Incentives play an important role in motivating the interest of voluntary participation, even more importantly a rewards scheme through which to determine how rewards are earned. Any future research would be to understand the necessary utility considerations for designing such a rewards incentive scheme that can adaptively incentivise agents for high quality participation.
\item From the agent behaviours observed when provided facilitation type incentives e.g. mobile data credit, the time lag between issuing the incentive and broadcasting the subject for the next surveillance task should be minimised to not more than a day. This recommendation applies to crowdsourcing schemes where the duration of 1 week is still considered near real-time.
\item Under this work we register the importance of training. For crowds that are mainly rural based, there is need for regular retraining and retooling programs. These could be a combination of online training via phone and onground live training sessions. 
\item A communication strategy is key to such a framework. Particularly a strategy that emphasizes feedback to the crowd was found to be of critical importance for this type of work. 
\item We found that social dynamics should be considered when implementing this type of system. We found not all participants were affected or responded the same towards certain incentives. While we did not do a rigorous study on this, we found that such dynamics as gender, age, social status, etc had a huge influence on the performance of the participants within this system.
\end{enumerate}

\section{Contributions}
This work sought to find the most optimal setup possible to crowdsource all year round real-time surveillance data from volunteer agents in disparate rural agricultural localities around the country, and eventually to understand what types of incentives would be effective in motivating such a crowd. Whilst the pilot was not explicitly set up to measure the incentive mechanism, we still observe some contributions from implementing this and tweaking incentives over 76 weeks. We summarise the contributions of this work as follows.
\begin{enumerate}
\item We present a crowdsourcing based system to address a real world problem in Uganda. We provide the requisite evidence for using the said system to actually collect over 7000 reports from a crowd consisting of four categories of people. The contribution is in how to set up such a system as well as the experiences in getting this working.
\item We also present a live example of how manipulating different incentives can affect the outcomes of a crowdsourcing task such as this. Particularly we talk about differently motivated intrinsic and extrinsic incentives and the effect on the crowd of varying them. However working on a real problem with real people presents its unique challenges, for example having a device stolen or having the technology fail can not be solved by improving the incentive mechanism.
\item For people particularly interested in this particular problem of crowdsourcing crop health surveillance data perhaps for other crops apart from cassava, we provide an analysis of the different categories of the \emph{crowd} and how they respond to different motivations. For example the more experienced members of the crowd respond differently to different types of incentives. Overall the social dynamics of the crowd in their localities also plays a big role, even though we were not able to demonstrate that concretely in this work.
\end{enumerate}
% One, the pilot equipped several small holder model farmers, experts and extension service agents with mobile smartphones and a surveillance app, able to report in-field observations on subjects of interest to the National Crops Resources and Research Institute(NaCRRI) in Uganda. Their contributions are mapped in real-time to a live interactive online map.\\
% Two, an early gain from this work is that the ad hoc surveillance crowdsourcing platform design presented here, has been deployed by the Cassava Regional Centre of Excellence at NaCRRI to monitor the health of cassava crop in regions where they are having cassava seed multiplication sites and farmers.\\
% Three, the work presents key insights in the participation behaviours of each of the crowd groups when exposed to several incentives over time. The insights are currently being used as a guide for engagement and incentive policies on an on-going scale-up project looking to crowdsource surveillance data from a network of over 200 small-holder farmers around Uganda.\\
% Four, the pilot managed to collect over 7000 images, which are useful for automated disease diagnosis, automated pest and symptom measurement, spatial-temporal modelling of pest and disease at NaCRRI, which however are not covered in the scope of this paper.

\subsection{Limitations of study}
The focus of this work was to understand how agricultural actors participate in a mobile crowdsourcing setup for surveillance of cassava viral disease and pests, by drawing insights from observations on reporting patterns of individuals contributors, behaviours of  their sub-groupings, their response to different incentives and the effectiveness of training exercises for rural crowds. It also looked at early benefits such a system presents in bridging the current gaps in monitoring of crop health. 

While we provide insights based on 76 weeks of piloting this system, we did not set it up systematically at the onset as a controlled experiment, and as such the conclusions from this work need to be looked at in that light. However it is also important to note that it is very hard to have a controlled experiment for an actual problem like this where the actors are playing in a rural field that we have no control over. Things like stolen phones, the actors falling sick, or getting involved in other activities do affect the outputs of such a system and it would be hard to tie them to the incentive mechanism for example.

\section{Conclusion}
 
This work harnessed the ubiquity of farmer communities and crowds using smartphones, to provide all year-round near real-time surveillance and monitoring data of cassava viral diseases and pests in Uganda as a supplement to the standard cassava crop surveys carried out annually. We obtained 7000 reports which is about one third of the expected number of reports if the system were operating in a very controlled environment with no phones stolen, no technical difficulties, and everyone motivated to send in data over 76 weeks. The goal is to eventually use this data to build an automated diagnostic tool for cassava diseases as well as use the data to provide a real time situation map of the state of disease in the whole country. Several issues come out of this work that require further investigation for example the social dynamics of this particular crowd, how to control the quality of images uploaded in such a system and how to incentivize the different categories of the crowd. This will form the substance of our future work. 

% With massive regular real-time reports, it is possible to forecast the disease risk burden for agricultural regions across the country with spatial-temporal modelling.\\
% We foresee the work here contributing many important results towards the goal of creating sustainable livelihoods in agriculture for small-holder farmers, enhancing precision surveillance under budget constrains and of timely interventions to curb viral crop disease and pest spread.
%\subsubsection{Acknowledgments.}

% This research work is done under the Artificial Intelligence \& Data Science Research Labs, www.air.ug , on the PEARL initiative with 
%funding from the Bill \& Melinda Gates Foundation.
\section{Acknowledgement}
We would like to acknowledge our partners from the government of Uganda, the National Crops Resources Research Institute (NaCRRI) and our funders, the Bill and Melinda Gates foundation (grant No. OPP1112548). 
% under whose auspices we did this work and who are actively engaged in the piloting of this system to support their surveillance functions. This work was funded by a grant from the Bill and Melinda Gates foundation, grant No. OPP1112548.

\bibliographystyle{aaai}
\bibliography{references}

\begin{thebibliography}{}

\bibitem[\protect\citeauthoryear{Agapie, Teevan, and
  Monroy-Hern{\'a}ndez}{2015}]{agapie2015crowdsourcing}
Agapie, E.; Teevan, J.; and Monroy-Hern{\'a}ndez, A.
\newblock 2015.
\newblock Crowdsourcing in the field: A case study using local crowds for event
  reporting.
\newblock In {\em Third AAAI Conference on Human Computation and
  Crowdsourcing}.

\bibitem[\protect\citeauthoryear{Chatzimilioudis \bgroup et al\mbox.\egroup
  }{2012}]{chatzimilioudis2012crowdsourcing}
Chatzimilioudis, G.; Konstantinidis, A.; Laoudias, C.; and Zeinalipour-Yazti,
  D.
\newblock 2012.
\newblock Crowdsourcing with smartphones.
\newblock {\em IEEE Internet Computing} 16(5):36--44.

\bibitem[\protect\citeauthoryear{Chklovski and
  Gil}{2005}]{chklovski2005towards}
Chklovski, T., and Gil, Y.
\newblock 2005.
\newblock Towards managing knowledge collection from volunteer contributors.
\newblock In {\em AAAI Spring Symposium: Knowledge Collection from Volunteer
  Contributors},  21--27.

\bibitem[\protect\citeauthoryear{Chklovski}{2003}]{chklovski2003learner}
Chklovski, T.
\newblock 2003.
\newblock Learner: a system for acquiring commonsense knowledge by analogy.
\newblock In {\em Proceedings of the 2nd international conference on Knowledge
  capture},  4--12.
\newblock ACM.

\bibitem[\protect\citeauthoryear{Haklay}{2013}]{haklay2013citizen}
Haklay, M.
\newblock 2013.
\newblock Citizen science and volunteered geographic information: Overview and
  typology of participation.
\newblock In {\em Crowdsourcing geographic knowledge}. Springer.
\newblock  105--122.

\bibitem[\protect\citeauthoryear{Kanefsky, Barlow, and
  Gulick}{2001}]{kanefsky2001can}
Kanefsky, B.; Barlow, N.~G.; and Gulick, V.~C.
\newblock 2001.
\newblock Can distributed volunteers accomplish massive data analysis tasks.
\newblock {\em Lunar and Planetary Science} 1.

\bibitem[\protect\citeauthoryear{Kittur \bgroup et al\mbox.\egroup
  }{2013}]{kittur2013future}
Kittur, A.; Nickerson, J.~V.; Bernstein, M.; Gerber, E.; Shaw, A.; Zimmerman,
  J.; Lease, M.; and Horton, J.
\newblock 2013.
\newblock The future of crowd work.
\newblock In {\em Proceedings of the 2013 conference on Computer supported
  cooperative work},  1301--1318.
\newblock ACM.

\bibitem[\protect\citeauthoryear{Mohan, Padmanabhan, and
  Ramjee}{2008}]{mohan2008nericell}
Mohan, P.; Padmanabhan, V.~N.; and Ramjee, R.
\newblock 2008.
\newblock Nericell: rich monitoring of road and traffic conditions using mobile
  smartphones.
\newblock In {\em Proceedings of the 6th ACM conference on Embedded network
  sensor systems},  323--336.
\newblock ACM.

\bibitem[\protect\citeauthoryear{Nuwamanya \bgroup et al\mbox.\egroup
  }{2015}]{ephraim2015}
Nuwamanya, E.; Baguma, Y.; Atwijukire, E.; Acheng, S.; and Alicai, T.
\newblock 2015.
\newblock Competitive commercial agriculture in sub saharan africa.
\newblock {\em International Journal of Plant Physiology and Biochemistry}
  7(2):12--22.

\bibitem[\protect\citeauthoryear{Otim-Nape \bgroup et al\mbox.\egroup
  }{2000}]{otim2000current}
Otim-Nape, G.; Bua, A.; Thresh, J.; Baguma, Y.; Ogwal, S.; Ssemakula, G.;
  Acola, G.; Byabakama, B.; et~al.
\newblock 2000.
\newblock The current pandemic of cassava mosaic virus disease in east africa
  and its control.
\newblock {\em The current pandemic of cassava mosaic virus disease in East
  Africa and its control.}

\bibitem[\protect\citeauthoryear{Paolacci, Chandler, and
  Ipeirotis}{2010}]{paolacci2010running}
Paolacci, G.; Chandler, J.; and Ipeirotis, P.~G.
\newblock 2010.
\newblock Running experiments on amazon mechanical turk.

\bibitem[\protect\citeauthoryear{Quinn and Bederson}{2011}]{quinn2011human}
Quinn, A.~J., and Bederson, B.~B.
\newblock 2011.
\newblock Human computation: a survey and taxonomy of a growing field.
\newblock In {\em Proceedings of the SIGCHI conference on human factors in
  computing systems},  1403--1412.
\newblock ACM.

\bibitem[\protect\citeauthoryear{Rana \bgroup et al\mbox.\egroup
  }{2010}]{rana2010ear}
Rana, R.~K.; Chou, C.~T.; Kanhere, S.~S.; Bulusu, N.; and Hu, W.
\newblock 2010.
\newblock Ear-phone: an end-to-end participatory urban noise mapping system.
\newblock In {\em Proceedings of the 9th ACM/IEEE International Conference on
  Information Processing in Sensor Networks},  105--116.
\newblock ACM.

\bibitem[\protect\citeauthoryear{Silvertown}{2009}]{silvertown2009new}
Silvertown, J.
\newblock 2009.
\newblock A new dawn for citizen science.
\newblock {\em Trends in ecology \& evolution} 24(9):467--471.

\bibitem[\protect\citeauthoryear{Singh \bgroup et al\mbox.\egroup
  }{2002}]{singh2002open}
Singh, P.; Lin, T.; Mueller, E.~T.; Lim, G.; Perkins, T.; and Zhu, W.~L.
\newblock 2002.
\newblock Open mind common sense: Knowledge acquisition from the general
  public.
\newblock In {\em OTM Confederated International Conferences" On the Move to
  Meaningful Internet Systems"},  1223--1237.
\newblock Springer.

\bibitem[\protect\citeauthoryear{Ssekibuule, Quinn, and
  Leyton-Brown}{2013}]{ssekibuule2013mobile}
Ssekibuule, R.; Quinn, J.~A.; and Leyton-Brown, K.
\newblock 2013.
\newblock A mobile market for agricultural trade in uganda.
\newblock In {\em Proceedings of the 4th Annual Symposium on Computing for
  Development}, ~9.
\newblock ACM.

\bibitem[\protect\citeauthoryear{Stevens and
  D’Hondt}{2010}]{stevens2010crowdsourcing}
Stevens, M., and D’Hondt, E.
\newblock 2010.
\newblock Crowdsourcing of pollution data using smartphones.
\newblock In {\em Workshop on Ubiquitous Crowdsourcing}.

\bibitem[\protect\citeauthoryear{Thiagarajan \bgroup et al\mbox.\egroup
  }{2009}]{thiagarajan2009vtrack}
Thiagarajan, A.; Ravindranath, L.; LaCurts, K.; Madden, S.; Balakrishnan, H.;
  Toledo, S.; and Eriksson, J.
\newblock 2009.
\newblock Vtrack: accurate, energy-aware road traffic delay estimation using
  mobile phones.
\newblock In {\em Proceedings of the 7th ACM conference on embedded networked
  sensor systems},  85--98.
\newblock ACM.

\bibitem[\protect\citeauthoryear{Van~Pelt and Sorokin}{2012}]{van2012designing}
Van~Pelt, C., and Sorokin, A.
\newblock 2012.
\newblock Designing a scalable crowdsourcing platform.
\newblock In {\em Proceedings of the 2012 ACM SIGMOD International Conference
  on Management of Data},  765--766.
\newblock ACM.

\bibitem[\protect\citeauthoryear{Zhu, Kraut, and
  Kittur}{2012}]{zhu2012organizing}
Zhu, H.; Kraut, R.; and Kittur, A.
\newblock 2012.
\newblock Organizing without formal organization: group identification, goal
  setting and social modeling in directing online production.
\newblock In {\em Proceedings of the ACM 2012 conference on Computer Supported
  Cooperative Work},  935--944.
\newblock ACM.

\end{thebibliography}

\end{document}